\documentclass[sigconf,table]{acmart}
\usepackage{array}
\usepackage{mathtools}
\usepackage{balance} 

\usepackage{hhline}
\DeclarePairedDelimiter{\norm}{\lVert}{\rVert}

\newcolumntype{P}[1]{>{\centering\arraybackslash}p{#1}}
\AtBeginDocument{%
  \providecommand\BibTeX{{%
    \normalfont B\kern-0.5em{\scshape i\kern-0.25em b}\kern-0.8em\TeX}}}

\copyrightyear{2021} 
\acmYear{2021} 
\setcopyright{acmcopyright}\acmConference[MM '21]{Proceedings of the 29th ACM International Conference on Multimedia}{October 20--24, 2021}{Virtual Event, China}
\acmBooktitle{Proceedings of the 29th ACM International Conference on Multimedia (MM '21), October 20--24, 2021, Virtual Event, China}
\acmPrice{15.00}
\acmDOI{10.1145/3474085.3475405}
\acmISBN{978-1-4503-8651-7/21/10}

\begin{document}
\fancyhead{} 
\title{ReconVAT: A Semi-Supervised Automatic Music Transcription Framework for Low-Resource Real-World Data}



\author{Kin Wai Cheuk}\additionalaffiliation{Institute of High Performance Computing (IHPC), Agency for Science, Technology and Research (A*STAR)}
\author{Dorien Herremans}
\affiliation{%
  \institution{Information Systems Technology and Design\\Singapore University of Technology and Design}
  \streetaddress{}
  \city{}
  \state{}
  \country{Singapore}
  \postcode{0000}  
  \state{}
}
\email{{kinwai_cheuk, dorien_herremans}@mymail.sutd.edu.sg}


\author{Li Su}
\affiliation{%
  \institution{Institute of Information Science\\Academia Sinica}
  \institution{}
  \streetaddress{}
  \city{}
  \state{}
  \country{Taiwan}
  \postcode{0000}  
  \state{}
}
\email{lisu@iis.sinica.edu.tw}



\begin{abstract}
Most of the current supervised automatic music transcription (AMT) models lack the ability to generalize. This means that they have trouble transcribing real-world music recordings from diverse musical genres that are not presented in the labelled training data. In this paper, we propose a semi-supervised framework, ReconVAT, which solves this issue by leveraging the huge amount of available unlabelled music recordings. The proposed ReconVAT uses reconstruction loss and virtual adversarial training. When combined with existing U-net models for AMT, ReconVAT achieves competitive results on common benchmark datasets such as MAPS and MusicNet. For example, in the few-shot setting for the string part version of MusicNet, ReconVAT achieves F1-scores of 61.0\% and 41.6\% for the note-wise and note-with-offset-wise metrics respectively, which translates into an improvement of 22.2\% and 62.5\% compared to the supervised baseline model. 
Our proposed framework also demonstrates the potential of continual learning on new data, which could be useful in real-world applications whereby new data is constantly available.
\end{abstract}

\ccsdesc[500]{Applied computing~Sound and music computing}
\ccsdesc[500]{Computing methodologies~Semi-supervised learning settings}
\ccsdesc[500]{Computing methodologies~Neural networks}

\keywords{semi-supervised training, virtual adversarial training, audio processing, automatic music transcription, music information retrieval}

\maketitle
\section{Introduction}
Automatic Music Transcription (AMT), is a fundamental problem in the field of Music Information Retrieval (MIR). According to the definition from the field of Music Information Retrieval (MIR)~\cite{Benetos2019AutomaticMT}, AMT aims at transcribing music audio files into symbolic representations such as piano rolls~\cite{cemgil2004bayesian} or music scores~\cite{carvalho2017towards, roman2018end, roman2019holistic}, which is very similar to Automatic Speech Recognition (ASR)~\cite{amodei2016deep, Chan2016ListenAA}. These symbolic representations have a wide range of applications including music indexing~\cite{cuthbert2010music21, sun2017query}, music generation~\cite{huang2020pop}, music recommendation system (MRS)~\cite{chen2001music, merono2017midi, yamaguchi2019music}, music analysis~\cite{kong2020large, jiang2019melody, liumei2021k}, and automatic music accompaniment~\cite{magalhaeschordify}.

Recent advances in fully supervised deep learning have enabled AMT models~\cite{Hawthorne2017OnsetsAF, hawthorne2018enabling, kim2019adversarial, kelz2019deep} to achieve state-of-the-art performance for solo piano pieces, given sufficient labelled training data. While acoustic audio recordings as well as the aligned midi labels for piano music can be easily obtained by using a hybrid acoustic/midi piano such as the Yamaha Disklavier~\cite{emiya2010maps}, this is not the case for other musical instruments such as violin and clarinet. At the time of writing, hybrid versions of these musical instruments are still not available. They are either midi controllers that lack the capability to produce original acoustic sound, or fully acoustic instruments without the capability to capture the real-time midi performance. Therefore, the paired acoustic and midi recordings for these instruments are very expensive to obtain, and hence, very limited. Supervised models fail to function well for these instruments.

Self-supervised or semi-supervised learning is an underexplored area in AMT. Existing unsupervised models have only been applied to specific musical instruments. For example, \citet{berg2014unsupervised} proposed an unsupervised graphical model using prerecorded key-wise piano samples to reconstruct the original signal. Upon successful reconstruction, the model could infer the transcription result via both the onset locations and the piano samples for reconstructing the spectrogram. \citet{DBLP:conf/ismir/ChoiC19} also employed a similar approach in their unsupervised drum transcription model. This approach, however, only works when the musical instrument is a percussion or plucked instrument type with a clear transient, immediately followed by a natural decay (piano is considered as a percussion instrument due to the hammering mechanism). Musical instruments which produce an increasing or fluctuating amplitude after the transient are unable to be represented by a fixed audio sample, and hence can not be properly transcribed using the above-mentioned approach. Examples of such instruments include string instruments that are capable of starting a long note softly followed by a crescendo through gradually increasing the bow pressure; and woodwind instruments that can sustain a note as long as the player's lung capacity can handle. 

\begin{figure}[t!]
  \centering
  \includegraphics[width=\linewidth]{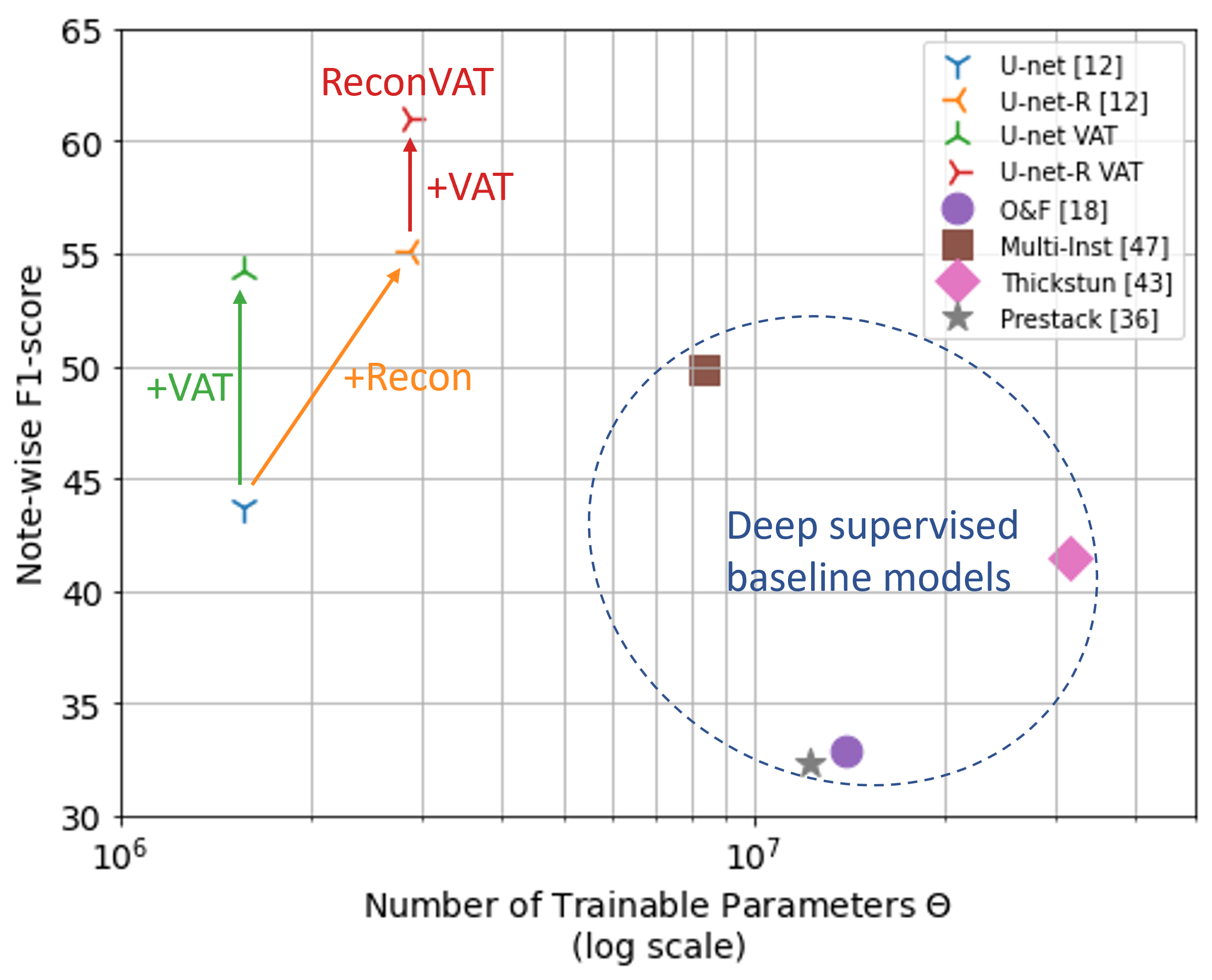}
  \caption{A scatter plot showing the note-wise F1-score and the number of model parameters for different models trained on the string version of MusicNet (see Section~\ref{sec:model_compactness} for details).}
  \Description{A scatter plot showing the note-wise F1-score  and the number of model parameters for different models trained on the string version of MusicNet.}
  \label{fig:model_compactness}
\end{figure}

In this paper, we propose a semi-supervised AMT framework based on virtual adversarial training (VAT) that leverages unlabelled data to improve the transcription accuracy with only a limited amount of labelled data. We integrate the spectrogram reconstruction into this framework, which we refer to as \textbf{ReconVAT}. ReconVAT works well on various musical instruments such as piano, string instruments, as well as woodwind instruments. We also show that our framework has important applications such as continual learning with new unlabelled recordings, and being able to transcribe music genres that are outside of the labelled training set. More importantly, all of this can be achieved with only a small number of model parameters compared to existing deep learning models for AMT as shown in Figure~\ref{fig:model_compactness}. This makes our framework attractive and practically usable for real-world applications deployed on mobile devices. To the best of our knowledge, this is the first semi-supervised deep learning framework for instrument-agnostic AMT at the time of writing.

The contributions of this paper can be summarized as follows:
\begin{itemize}
  \item We propose a semi-supervised framework for AMT that generalizes well across \emph{different kinds of musical instruments}.
  \item We leverage existing models by integrating them into the proposed semi-supervised framework to achieve state-of-the-art transcription accuracy for low-resource scenario.
  \item We demonstrate possible applications in continual learning on music genres that are not present in the train set.
\end{itemize}

\begin{figure*}[ht]
  \centering
  \includegraphics[width=\linewidth]{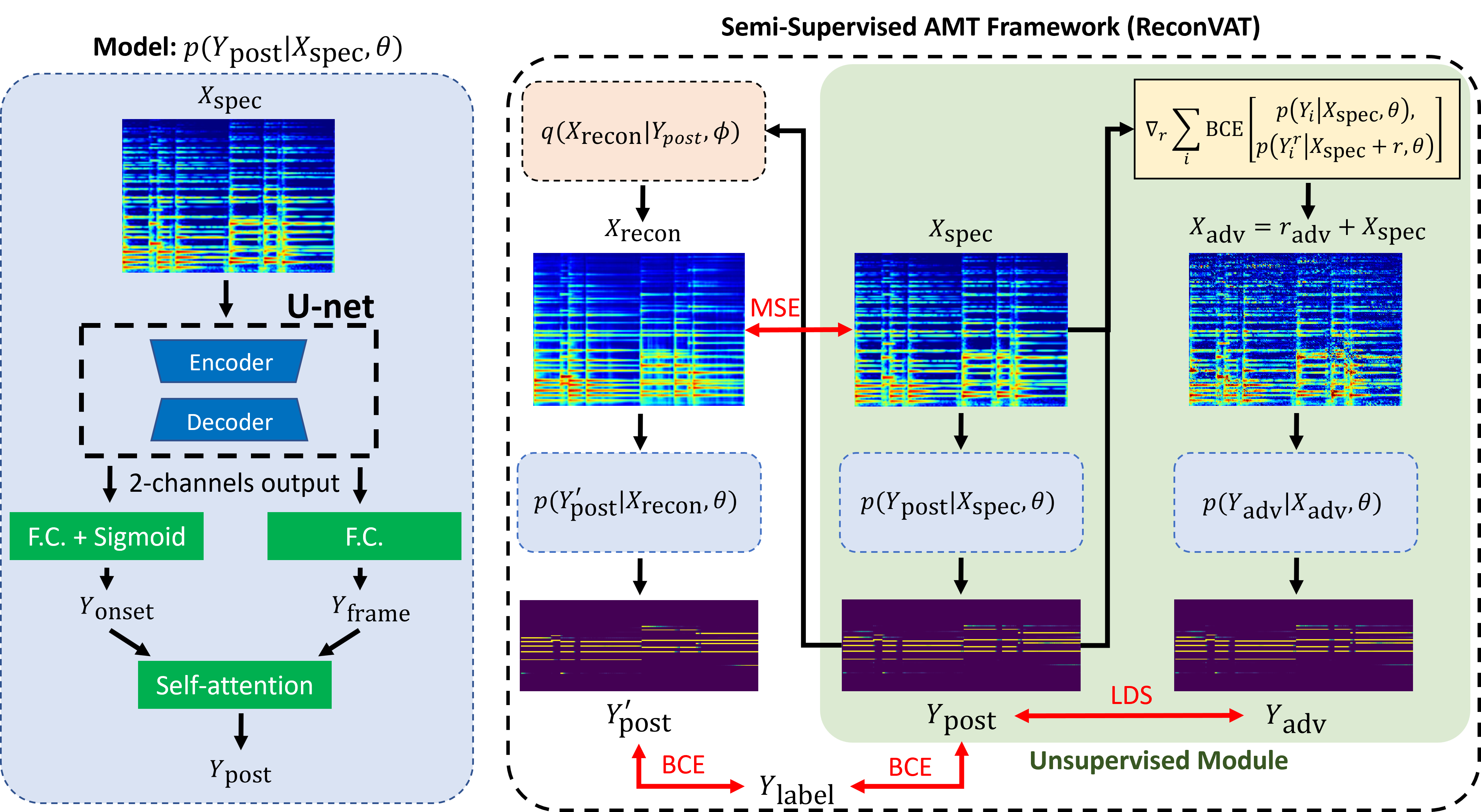}
  \caption{The left-hand side of the figure shows the modified version of the AMT model proposed in~\cite{cheukICPR} such that it supports onset prediction. The right-hand side of the figure shows the proposed framework for semi-supervised AMT. For simplicity, we omit the onset prediction in the figure by showing the case when $i=\{\text{post}\}$. The region highlighted in green is the unsupervised module which supports training not only on labelled samples, but also on unlabelled samples.}
  \Description{ReconVAT, our proposed semi-supervised framework for AMT.}
  \label{fig:model}  
\end{figure*}

\section{Method}
In this section, we will formulate automatic music transcription (AMT) mathematically. Then we will introduce related work, and describe how to combine both spectrogram reconstruction~\cite{cheukICPR} and VAT~\cite{VAT_semi} to be our proposed semi-supervised framework ReconVAT.

\subsection{Problem Definition}
The goal of AMT is to convert audio data into symbolic music data~\cite{Benetos2013AutomaticMT,Benetos2019AutomaticMT}. In this paper, we consider the case of converting spectrograms into piano rolls. Given an input spectrogram $X_\text{spec} \in [0,1]^{T\times F}$, where $T$ is the number of timesteps and $F$ is the number of frequency bins, we want to have a model $p(Y_\text{post}|X_\text{spec},\theta)$, with a set of trainable parameters~$\theta$, that infers the posteriorgram $Y_\text{post}\in [0,1]^{T\times N}$. Here $N$ is the note range for the musical instrument, for example, $N=88$ for piano transcription since there are 88 keys on the keyboard. The ground truth piano roll $Y_\text{roll}\in \{0,1\}^{T\times N}$ is the symbolic notation we want to predict. This is done by simply applying a threshold (e.g. $0.5$) to the $Y_\text{post}$.

\subsection{Spectrogram Reconstruction}
\label{sec:spec_recon}
\sloppy In \citet{cheukICPR}, the authors proposed a model consisting of a transcriber $p(Y_\text{post}|X_\text{spec},\theta)$ and a reconstructor $q(X_\text{recon}|Y_\text{post},\phi)$. The reconstructor uses the posteriorgram generated from the transcriber as input to reconstruct the spectrograms $X_\text{recon}$. Therefore, in addition to the transcription loss $L_\text{trans}(Y_\text{post}, Y_\text{label})$, there is also a reconstruction $L_\text{recon}(X_\text{recon}, X_\text{spec})$ to be minimized.

The reconstructed spectrograms $X_\text{recon}$ are then used to train the same transcriber $p(Y'_\text{post}|X_\text{recon},\theta)$ again, resulting in one extra transcription loss $L_\text{trans}(Y'_\text{post}, Y_\text{label})$. \citet{cheukICPR} has shown that training the model in this manner results in a consistently better model. Their reported results, however, are unable to beat the state-of-the-art AMT models. We will show in Table~\ref{tab:full_MAPS} that their model can be modified to compete with the state-of-the-art AMT models. Although it is not demonstrated in their paper, they also claim that their model has the potential to be trained in an unsupervised manner. We will therefore also show in Section~\ref{sec:proposed_framework} that when combined with virtual adversarial training~\cite{VAT_semi}, we can modify the spectrogram reconstruction framework~\cite{cheukICPR} to be a semi-supervised model for AMT.

\subsection{Virtual Adversarial Training}
Virtual adversarial training (VAT), as presented by ~\citet{VAT_semi} is an extended version of adversarial training (AT) proposed by~\citet{adversarial}. In AT, labels are required to calculate the the adversarial vectors. In the case where we do not have access to the labels, \citet{VAT_semi} proved that the adversarial vector can be obtained via Equation~(\ref{eq:VAT}):



\begin{equation}
\label{eq:VAT}
r^\text{VAT}_\text{adv} = \epsilon\left(\nabla_r D\left[p(Y_\text{pred}|X, \theta), p(Y_\text{adv}|X+r, \hat{\theta})\right] \right)\,,
\end{equation}
\noindent where $r$ is a randomly initialized noise vector, and $Y_\text{adv}$ is the output obtained using the adversarial input $X_\text{adv} = X + r$.

By doing so, it is possible to perform adversarial training using unlabelled datasets. Most existing literature applies VAT to static classifications~\cite{amodei2016deep,Chan2016ListenAA,park2019adversarial, miyato2016adversarial, kuwahara2019model, kreyssig2020cosine}. While SeqVAT~\cite{chen2020seqvat} is designed for sequential labelling, it is a one-hot prediction system. To the best of our knowledge, ReconVAT is the first framework capable of multi-hot sequential labelling for polyphonic AMT.
In the next section, we will describe how to combine both spectrogram reconstruction and VAT to obtain a semi-supervised framework for AMT.

\subsection{Proposed Framework -- ReconVAT}
\label{sec:proposed_framework}

The left-hand side of Figure~\ref{fig:model} shows our modified version of the model proposed by~\citet{cheukICPR}. We improve it by introducing a two-channel output, one channel for the onset prediction $Y_\text{onset}$, and another channel for the frame feature extraction $Y_\text{frame}$. The posteriorgram is obtained from a self-attention layer which takes the concatenation of $Y_\text{onset}$ and $Y_\text{frame}$ as the input. This modification increases the model's flexibility. For example, if we want to also include the offset prediction $Y_\text{offset}$, we can have a three-channel output instead. For simplicity, we will only explore the case of one-channel (without onset prediction) and two-channel (with onset prediction) prediction in this paper. The implementation details will be discussed in Section~\ref{sec:implementation}.

The right-hand side of Figure~\ref{fig:model} shows our proposed ReconVAT. It consists of three branches. The framework starts with the middle branch where it takes $X_\text{spec}$ as the input and outputs a posteriorgram $Y_\text{post}$. The branch on the left then takes the $Y_\text{post}$ as its input and generates a reconstructed spectrogram $X_\text{recon}$ using the reconstructor $q(X_\text{recon}|Y_\text{post},\phi)$ mentioned in~\ref{sec:spec_recon}. The reconstructed spectrogram $X_\text{recon}$ is passed to the same model again to obtain another posteriorgram $Y'_\text{post}$. The two posteriorgrams $Y_\text{post}$ and $Y'_\text{post}$ should be as close to the label $Y_\text{label}$ as possible. 

The branch on the right-hand side is the unsupervised module which uses VAT. To obtain the adversarial spectrogram $X_\text{adv}$, we apply a modified version of VAT that works better for AMT (Section~\ref{sec:implementation}). Using this adversarial spectrogram, we obtain another posteriorgram $Y_\text{post}$ via the same model.

For labelled spectrograms, all three branches are used. For unlabelled spectrograms, we only use the middle and the right branches (highlighted in green in Figure~\ref{fig:model}). By doing so, this allows us to train our model with unlabelled data. This framework is trained by minimizing both the supervised loss $L_\text{l}$ (Equation~\ref{eq:L_l}) and the unsupervised loss $L_\text{ul}$ which will be discussed in detail in Section~\ref{sec:objective}.

\vspace{-1.5mm}
\section{Experiments}
\label{sec:experiments}

In this section, we describe the datasets and the experiments for demonstrating the power of our proposed semi-supervised framework ReconVAT.

\subsection{MAPS dataset}

The MAPS dataset~\cite{emiya2010maps} consists of nine folders, each folder contains 30 full-length midi recordings. In seven of these folders, the audio recordings are synthesized from the midi annotations using different virtual piano software such as Steinberg, Native Instruments, and Sampletekk. Only in the folders \texttt{ENSTDkAm} and \texttt{ENSTDkCl}, the audio recordings are recorded simultaneously with the midi recordings using a Yamaha Disklavier. We follow the existing consensus~\cite{Hawthorne2017OnsetsAF,Sigtia2015AnEN,cheukICPR, kelz2019deep, pedersoli2020improving} that the seven folders containing artificially generated audio recordings should be used as the training set, and the other two folders, \texttt{ENSTDkAm} and \texttt{ENSTDkCl}, as the test set.

Since some music pieces appear in both the training and the test set, we follow the existing literature~\cite{Sigtia2015AnEN,Hawthorne2017OnsetsAF} to remove overlapping songs from the training set that are also present in the test set, thus reducing the size of the training set from 210 music pieces down to 139 pieces.
Following existing conventions~\cite{Sigtia2015AnEN,Hawthorne2017OnsetsAF, hawthorne2018enabling, cheukICPR}, all audio recordings are downsampled from $44.1$~kHz to $16$~kHz.

To demonstrate the effectiveness of our VAT model, we train our model using the following three versions of the MAPS dataset:
\subsubsection{Full version}
This version uses all 139 available pieces from MAPS as the labelled training set. To demonstrate the ability of leveraging unlabelled data using our VAT model, we use the training set from MAESTRO~\cite{hawthorne2018enabling} as the unlabelled dataset (967 music recordings). The labelled training batch size $N_\text{l}$ and the unlabelled training batch size $N_\text{ul}$ are both 8.

\subsubsection{Small version}
In this version, only one folder (\texttt{AkPnBcht}, containing 23 non-overlapping songs) from MAPS is used as the labelled training set. We keep using the same 967 music recordings from MAESTRO as our unlabelled set for our VAT model. Again, $N_\text{l}$ and $N_\text{ul}$ are both 8 in this version.

\subsubsection{One-shot version}
Only one music recording (\texttt{chp\_op31} from the \texttt{AkPnBcht} folder) is used as the labelled training set. The unlabelled set consists of the same 967 music recordings from MAESTRO as the above two versions. Due to the fact that there is only one labelled training sample, $N_\text{l}$ is 1 and the unlabelled training batch size $N_\text{ul}$ remains 8.

\subsection{MusicNet dataset}

 MusicNet~\cite{Thickstun2016LearningFO} contains both audio recordings and annotations of various types of musical instruments such as those from the string family and the woodwind family. To prove that our model also works for different types of musical instruments, we perform our experiments on the following variations of MusicNet:

\subsubsection{String version}
\label{sec:string_musicnet}
In the official training set provided by MusicNet, there are 8 genres of music that contain string instruments. We select only one piece from each genre from the official training set, forming our own labelled training set. The remaining pieces of each genre are used as the unlabelled training set for our VAT framework. By doing so, there are eight labelled samples and 104 unlabelled samples in our training set. We pick four string pieces from the official test set provided by MusicNet as our test set. The details of data splitting can be found in the supplementary material\footnotemark[2].
The labelled training batch size $N_\text{l}$ and the unlabelled training batch size $N_\text{ul}$ are both 8.

\subsubsection{Woodwind version}
\label{sec:woodwind_musicnet}
Similar to the string version, we pick only one piece from six different woodwind genres from MusicNet as the labelled training set and use the remaining pieces in each genre as the unlabelled training set. This results in six labelled training samples and 21 unlabelled training samples. The official test set provided by MusicNet contains only two pieces (1819, 2416) from the woodwind family, which belong to the Pairs Clarinet-Horn-Bassoon genre. We use these two pieces as our test set. Again, more details can be found supplementary material\footnotemark[2].
$N_\text{l}$ is 1 and $N_\text{ul}$ is 8 in this version.

\subsection{Data Processing}
We extract Mel spectrograms on-the-fly from the audio clips using a GPU-based audio processing library nnAudio~\cite{cheuk2019nnaudio}. Following~\citet{Hawthorne2017OnsetsAF}, we use a Hann window size of 2,048, a hop size of 512, and 229 Mel bins as the parameters of our Mel spectrograms $X_\text{spec}$. To extract a fixed length spectrogram, we crop the audio clips into segments of 327,680 sample points using random sampling during each iteration, which results in Mel spectrogram with 640 timesteps, and 229 Mel frequency bins. We compress the magnitude of the spectrograms by taking the natural logarithm and then normalizing the magnitude for each spectrogram into the range $[0,1]$. i.e. $X_\text{spec} \in [0,1]^{640\times 229}$.

As for our ground truth labels, we extract the onset, duration, and pitch information from the midi annotations to produce tsv files for the ground truth. These tsv files are read and converted into piano rolls in the form of a binary matrix $Y_\text{label}\in \{0,1\}^{640, F}$. Since most musical instruments in the dataset are within the 88 notes range (note A0 to note C8), we use $F=88$ in all our experiments.

\subsection{Implementation Details}
\label{sec:implementation}
All models and experiments, including the baseline models, are implemented in PyTorch. To ensure transparency and fairness, we train all our models without tricks such as label smoothing~\cite{MultiInstrument}, weighted cross entropy~\cite{Hawthorne2017OnsetsAF}, and focal loss~\cite{Wu2019focalAMT, lin2017focal}. We believe that these tricks would in general improve the the transcription accuracy, and it is beyond the scope of this paper to explore this.

We adopt U-net models specifically designed for pitch detection~\cite{cheukICPR,Hung2019MusicalCS} and integrate them into the VAT framework~\cite{VAT_semi}. While we follow  mostly the same design as in~\cite{cheukICPR}, we modify the final layer of the decoder so that it has the flexibility to output two channels as shown in Figure~\ref{fig:model}. One of the channels is fed to a fully connected layer with sigmoid activation to predict the onsets $Y_{\text{onset}}\in [0,1]^{T\times F}$, and the other channel is fed to a linear fully connected layer to obtain the features $Y_{\text{frame}}\in \mathbb{R}^{T\times F}$. The concatenated output $Y_{\text{onset}} \bigoplus Y_{\text{frame}}$ is fed to a relative local 1D self-attention layer~\cite{self-attention_shaw2018,standalone_Ramachandran} to obtain the posteriorgram $Y_{\text{post}}\in [0,1]^{T\times 88}$. We binarize the posteriorgram with a threshold of $0.5$ to obtain the predicted piano roll $Y_{\text{roll}}\in \{0,1\}^{T\times 88}$. If the two-channel output is used, we follow the inference method from the Onsets and Frames model~\cite{Hawthorne2017OnsetsAF} to obtain a refined piano roll by using both $Y_{\text{onset}}$ and $Y_{\text{frame}}$ to filter out notes that do not have a onset. Otherwise, we directly use the posteriorgram to obtain the piano roll. In addition, we also replace all of the LSTM layers in~\cite{cheukICPR} with local relative self-attention layers, since it has been shown that self-attention layers perform as good as LSTM layers while providing the extra benefit of being able to train in parallel~\cite{won2019interpretable, MultiInstrument}. 

We also modify the original VAT method~\cite{VAT_semi} so that it works better for AMT. Firstly, since polyphonic AMT is a timestep-wise multiclass classification problem (multiple pitches can occur at the same time), we replace the Kullback–Leibler divergence (KL-div) with binary cross entropy (BCE) when calculating the local distributional smoothness (LDS). Secondly, we normalise the adversarial vector $r_{\text{adv}}$ along the timestep dimension as shown in Equation~(\ref{eq:r_adv}):

\begin{equation}
\label{eq:r_adv}
  r_{\text{adv}} = \epsilon \begin{bmatrix}\frac{g_1}{{\norm{g_1}}_2}, \frac{g_2}{{\norm{g_2}}_2}, \cdots, \frac{g_T}{{\norm{g_T}}_2} \end{bmatrix} 
\end{equation}

\noindent where $\epsilon$ is a parameter that controls the magnitude of the adversarial vector $r_{\text{adv}}$, and $g_t$ for $1\leq t \leq T$ is the timestep-wise gradient obtained from Equation~(\ref{eq:VAT gradient})
\begin{equation}
\label{eq:VAT gradient}
g=\nabla_r \sum_{i}\text{BCE}\left[p(Y_{\text{i}}|X_{\text{spec}},\hat{\theta}), p(Y_{\text{adv}}|X_{\text{spec}}+r,\hat{\theta}) \right]\,.
\end{equation}

If the onsets prediction module is included, then $i=\{\text{onset}, \text{post}\}$. Otherwise, there is only one term in Equation~(\ref{eq:VAT gradient}), i.e. $i=\{\text{post}\}$. As in~\cite{VAT_semi}, the weight of the model is considered as a constant $\hat{\theta}$ when calculating the gradient $g$.

Once we obtain the adversarial vector $r_{\text{adv}}$, we can calculate the LDS. By the same logic as above, the LDS can contain either one or two terms depending on the model output:
\begin{equation}
\label{eq:LDS}
\text{LDS}_*= \frac{\sum_{i}\text{BCE}\left[p(Y_{\text{i}}|X^{*}_{\text{spec}},\hat{\theta}), p(Y_{\text{adv}}|X^{*}_{\text{spec}}+r_{\text{adv}},\theta) \right]}{N_*}\,.
\end{equation}

From Equation~\ref{eq:LDS}, we can see that the label $Y_i^{\text{label}}$ is not required to calculate the LDS. Therefore, LDS is an unsupervised loss that can be calculated using both labelled spectrograms $X^{\text{l}}_{\text{spec}}$ and unlabelled spectrograms $X^{\text{ul}}_{\text{spec}}$. We will denote the LDS calculated using $X^{\text{l}}_{\text{spec}}$ as $\text{LDS}_\text{l}$ and the LDS calculated using $X^{\text{ul}}_{\text{spec}}$ as $\text{LDS}_\text{ul}$. Unlike the original VAT~\cite{VAT_semi}, we normalise $\text{LDS}_\text{l}$ and $\text{LDS}_\text{ul}$ by its respective batch size $N_\text{l}$ and $N_\text{ul}$, rather than summing both $\text{LDS}_\text{l}$ and $\text{LDS}_\text{ul}$ together and normalize with $N_\text{l} + N_\text{ul}$. By doing so, we prevent $N_\text{ul}$ from interfering with $\text{LDS}_\text{l}$ and $N_\text{l}$ from interfering with $\text{LDS}_\text{ul}$.

\subsection{Training Objective and Optimization}
\label{sec:objective}
As mentioned in Section~\ref{sec:implementation}, we have the supervised objective $L_{\text{l}}$ that requires labels, and the unsupervised objective $L_{\text{ul}}$ that does not require any label. The final objective $L$ being minimized during training contains three terms as shown in Equation~(\ref{eq:objective}):
\begin{equation}
\label{eq:L_l}
L_{\text{l}}=\sum_{i}\text{BCE}\left[Y_i, Y^{\text{label}}_{i} \right] + \sum_{i}\text{BCE}\left[Y^{\text{recon}}_{i}, Y^{\text{label}}_{i} \right]
\end{equation}

\begin{equation}
\label{eq:L_ul}
L_{\text{ul}}=\frac{\text{LDS}_{\text{l}}+\text{LDS}_{\text{ul}}}{2}
\end{equation}

\begin{equation}
\label{eq:objective}
L = L_{\text{l}} + \alpha L_{\text{ul}} + L_\text{recon}
\end{equation}
where $\alpha$ is the weighting for $L_\text{ul}$, which is set to $1$ throughout all our experiments; $L_\text{recon}$ is the reconstruction loss mentioned in Section~\ref{sec:spec_recon}. We observe the same model behaviour as reported in~\cite{VAT_semi}, that is, controlling the $\epsilon$ in Equation~(\ref{eq:r_adv}) alone is sufficient to control the model performance without the need to change $\alpha$.

To minimize the objective $L$, we use Adam~\cite{kingma2014adam} optimizer with a learning rate of $0.001$ and a learning rate decay of $2\%$ every 1,000 iterations. When training, our framework includes three forward passes during each iteration. One forward pass for $L_\text{l}$, one forward pass for $\text{LDS}_\text{l}$, and one forward pass for $\text{LDS}_\text{ul}$. We define one epoch as 10 iterations. During the parameter search, we split our training set into 80\% for training and 20\% for validating. The optimal value for $\epsilon$ in Equation~(\ref{eq:r_adv} is mostly within the range between $1$ and $2$, and depends on the model architecture and the dataset. This value can be easily obtained after a few trials.

\subsection{Evaluation Metrics}
\label{sec:evaluation}
Following existing literature~\cite{cheukICPR, Cheuk_IJCNN2021, Hawthorne2017OnsetsAF, hawthorne2018enabling, kim2019adversarial, kelz2019deep}, we report the frame-wise, note-wise, and note-with-offset-wise metrics to evaluate our model performance comprehensively. For note-wise metric, we use a onset tolerance of 50ms; for note-with-offset-wise metric, we use an offset tolerance of 50ms or $20\%$ of the note duration, whichever is larger~\cite{bay2009evaluation}. Readers are referred to \citet{Cheuk_IJCNN2021} which explains the differences between these metrics in detail in their Section IV-C. In our experiments, we use the implementations from \texttt{mir\_eval}\footnote{\texttt{https://github.com/craffel/mir\_eval}} to calculate and report the above-mentioned metrics.



\section{Results}
\label{sec:results}

\subsection{Effectiveness of VAT}
We compare our proposed models to the Onsets and Frames model~\cite{Hawthorne2017OnsetsAF} and the Multi-Instrument AMT model~\cite{MultiInstrument} as they show good performance on the MAPS and MusicNet datasets respectively. We exclude the models proposed by \citet{pedersoli2020improving} and \citet{Thickstun2017InvariancesAD} in our results below since their performance is worse than the Multi-Instrument AMT model~\cite{MultiInstrument}. We use R to represent the reconstruction module, and O to represent the onset module. Therefore U-net-RO means that the U-net model contains both a reconstruction and onset module. The columns represent the precision (P), recall (R), and F1-score for each of the metrics mentioned in Section~\ref{sec:evaluation}. Our proposed models and the baseline models are trained on the same labelled data, and only the proposed semi-supervised models are able to leverage the unlabelled data mentioned in Section~\ref{sec:experiments}.
\subsubsection{Full MAPS}
\label{sec:result_full_MAPS}
We can see that when using the VAT (row A3-A4, A7, A8), all three metrics generally improve compared to their respective counterparts without the VAT (row A1-A2, A5, A6). When using onset inference (A5-A8), the note-wise and note-with-offset-wise metrics are improved by at least 7 percentage points.  The model using both the onset inference as well as our proposed framework (row A8) performs as good as the state-of-the-art Onsets and Frames model~\cite{Hawthorne2017OnsetsAF} (row 9) for this dataset.

\subsubsection{Small MAPS}
The middle part of Table~\ref{tab:full_MAPS} shows that when the number of labelled training samples is reduced by over $80\%$ from $139$ to $23$ audio clips, the advantage of the VAT module becomes more obvious. Similar for the full MAPS dataset, the models with VAT module outperform their counterparts that do not use VAT. Moreover, our proposed framework (row B8) outperforms the Onsets and Frames model (B9) by 6, 5.1, 4.4 percentage points in terms of frame-wise, note-wise, and note-with-offset-wise F1-scores, which can be translated into improvements in performance of $11.5\% $, $8.1\% $, and $14.1\% $ respectively. 

\subsubsection{One-shot MAPS}
The bottom part of Table~\ref{tab:full_MAPS} shows that when we reduce the number of labelled training audio clip even further to only one, our proposed framework (C8) outperforms the Onsets and frames model (C9) by 23.7, 17, and 12.9 percentage points.

Between the models that use and do not use onset inference, we can see that onset inference has the tendency of decreasing the frame-wise F1-score while improving the note-wise F1-scores. This is due to the unreliability of the frame-wise metric~\cite{Hawthorne2017OnsetsAF,Cheuk_IJCNN2021}. \citet{Cheuk_IJCNN2021} has provided a few examples and shown that a high frame-wise score does not guarantee a good transcription. Nonetheless, these three experiments have shown that VAT is a very effective semi-supervised method, that allows the use of unlabelled training samples to greatly improve the model performance in cases where the number of labelled samples is scarce.

\subsubsection{String MusicNet}
The top section of Table~\ref{tab:string_MusicNet} shows the performance of our proposed framework on the string subset of MusicNet~(\ref{sec:string_musicnet}). Interestingly, using the onset inference (row D1-4 and D9) does not improve the transcription accuracy in this setting, on the contrary, it worsens the model performance. Although most models with the VAT outperform their counterparts without the VAT, U-net-RO VAT on row B9 performs worse than its counterpart without the VAT. There are two possible reasons for this. First, we believe that the onset inference only works well for piano only, and it cannot generalize well to other musical instruments such as those from the string and the woodwind family. Second, we believe that the onset labels for MusicNet are not completely accurate, since the labels are generated using dynamic time warping (DTW)~\cite{Thickstun2017InvariancesAD}. Therefore, inaccurate onset labels might confuse the VAT. Using no labels might be better than using inaccurate labels, which is one of the advantages of using VAT.

Now, let us consider models (row D5-D8) that do not use onset inference. We will use the Multi-Instrument AMT model (row D10)~\cite{MultiInstrument}, which is the state-of-the-art model for the MusicNet dataset at the time of writing, as the baseline model. Since the baseline model~\cite{MultiInstrument} is much deeper than the U-net model (row D5), it outperforms the U-net model. By applying the reconstruction module to the U-net model (row D6), the U-net model begins to outperform the baseline model. When we further apply VAT to the U-net models (row D7-D8), the transcription accuracy becomes even better. The best model, U-net-R VAT (row D8), outperforms the baseline model by 3.9, 11.1, and 11.3 percentage points in terms of frame-wise, note-wise, and note-with-offset-wise metrics.

\subsubsection{Woodwind MusicNet}
The bottom section of Table~\ref{tab:string_MusicNet} shows the results for the woodwind subset of MusicNet~(Section~\ref{sec:woodwind_musicnet}). Since the Onsets and Frames model does not work well for this dataset either, we did not spend time experimenting with it. Just like all of the results reported above, the VAT module is very effective in improving the transcription accuracy. The best model being the one with both the reconstruction and the VAT module (row E4), and it outperforms the baseline model by $10.3$, $6.9$ percentage points in terms of note-wise and note-with-offset-wise metrics. The improvement for the frame-wise metrics is not obvious, however, we must keep in mind that this is not a reliable metric to evaluate the transcription accuracy as pointed out previously in Section~\ref{sec:evaluation} as well as existing literature~\cite{Cheuk_IJCNN2021,Hawthorne2017OnsetsAF}.

\subsection{Model Compactness}
\label{sec:model_compactness}

A comparison of number of trainable model parameters for the baseline models and the proposed models is shown in Figure~\ref{fig:model_compactness}. It can be seen from the figure that a deep model does not necessarily yield a high transcription accuracy when the labelled training data is limited. The Onsets and Frames model~\cite{Hawthorne2017OnsetsAF} and the Prestack-Unet~\cite{pedersoli2020improving} have a high number of parameters, yet they do not perform well when the labelled data is scarce. While Thickstun's model~\cite{Thickstun2017InvariancesAD} performs better than the two baseline models, its number of parameters is 10 times more than our proposed framework (U-net-R VAT). We use the Resnet-18 version of Prestack-Unet since the Resnet-32 version is too huge to run on our GPU.
Another baseline model, the Multi-Instrument AMT model, performs better than the plain U-net model. With VAT, however, the U-net models already outperform the baseline model while keeping the number of trainable parameters low. We can also see that the VAT improves the model performance without adding extra parameters to the model. Therefore, VAT is a very effective method to improve the transcription accuracy by leveraging unlabelled training data when the labelled training data is limited.

\begin{table*}[h]
\centering
 \caption{Transcription accuracy (mean + STD) in terms of different metrics and their respective precision (P), recall (R), and F1-score (F1) when training on different variations of the MAPS dataset. The accuracy values are averaged over the test clips in the dataset.}
\small
\resizebox{\textwidth}{!}
{\begin{tabular}{ll|ccc|ccc|ccc|}
\cline{3-11}
\multicolumn{1}{c}{}  &\multicolumn{1}{c|}{}                             & \multicolumn{3}{c|}{\textbf{Frame}} & \multicolumn{3}{c|}{\textbf{Note}}            & \multicolumn{3}{c|}{\textbf{Note w/ offset}}  \\ \cline{3-11} 
&\multicolumn{1}{c|}{\textbf{Full}}                           & P              & R    & F1          & P             & R             & F1            & P             & R             & F1            \\ \cline{2-11} 
A1&\multicolumn{1}{|l|}{U-net~\cite{cheukICPR}}   & 84.6 ± 6.0& 70.8 ± 8.9& 76.7 ± 6.5& 55.8 ± 12.6& 62.0 ± 11.9& 58.4 ± 11.7& 34.5 ± 11.3& 38.5 ± 12.0& 36.2 ± 11.4        \\ \cline{2-11} 
A2&\multicolumn{1}{|l|}{U-net-R~\cite{cheukICPR}} & 86.2 ± 6.2& 72.7 ± 10.0& 78.4 ± 7.0& 68.5 ± 10.5& 61.0 ± 13.1& 64.2 ± 11.4& 45.5 ± 11.1& 40.8 ± 12.9& 42.8 ± 11.9 \\ \cline{2-11} 
A3&\multicolumn{1}{|l|}{U-net VAT} & 86.6 ± 5.4& 71.5 ± 9.4& 77.9 ± 6.7& 64.5 ± 13.2& 64.2 ± 12.6& 64.0 ± 12.3& 40.8 ± 11.6& 40.9 ± 12.1& 40.6 ± 11.5 \\ \cline{2-11} 
A4&\multicolumn{1}{|l|}{U-net-R VAT} & 88.8 ± 6.0& \textbf{72.7 ± 9.0}& \textbf{79.5 ± 6.5}& 74.0 ± 9.3& 63.3 ± 13.3& 67.9 ± 11.2& 49.7 ± 10.1& 42.9 ± 12.5& 45.8 ± 11.3 \\ \cline{2-11}
A5&\multicolumn{1}{|l|}{U-net-O} & 89.6 ± 6.0& 58.8 ± 9.9& 70.4 ± 7.5& 85.8 ± 7.8& 66.3 ± 11.0& 74.5 ± 9.2& 53.1 ± 9.5& 41.5 ± 11.3& 46.4 ± 10.6 \\ \cline{2-11}
A6&\multicolumn{1}{|l|}{U-net-RO} & 89.9 ± 6.6& 60.4 ± 10.8& 71.6 ± 8.3& 86.1 ± 7.9& 67.3 ± 11.2& 75.2 ± 9.2& 52.8 ± 10.1& 41.7 ± 11.6& 46.4 ± 10.9 \\ \cline{2-11} 
A7&\multicolumn{1}{|l|}{U-net-O VAT} &\textbf{ 90.9 ± 6.1}& 60.5 ± 9.5& 72.2 ± 7.5& \textbf{89.8 ± 8.3}& 65.0 ± 11.1& 75.1 ± 9.5& \textbf{58.7 ± 9.9} & 42.9 ± 11.3& 49.4 ± 10.8 \\ \cline{2-11} 
A8&\multicolumn{1}{|l|}{U-net-RO VAT} & 85.9 ± 7.2& 72.0 ± 8.7& 77.9 ± 6.5& 80.9 ± 7.0& 70.6 ± 11.2& 75.1 ± 8.6& 54.3 ± 9.8& \textbf{47.6 ± 11.8}& \textbf{50.5 ± 10.6} \\ \cline{2-11} 
A9&\multicolumn{1}{|l|}{O\&F~\cite{Hawthorne2017OnsetsAF} } & 89.3 ± 6.4& 65.6 ± 9.7& 75.2 ± 7.3& 85.2 ± 7.8& \textbf{73.3 ± 11.4}& \textbf{78.6 ± 9.3}& 53.8 ± 9.8& 46.7 ± 12.0& 49.8 ± 10.9  \\ \cline{2-11}

&\multicolumn{1}{c|}{\textbf{Small}}& P& R& F1& P& R& F1& P& R& F1 \\ \cline{2-11}
B1&\multicolumn{1}{|l|}{U-net~\cite{cheukICPR}}   & 75.4 ± 6.6& 57.1 ± 9.6& 64.5 ± 7.3& 35.4 ± 8.3& 57.5 ± 11.6& 43.5 ± 8.9& 17.0 ± 6.9& 27.6 ± 10.6& 20.9 ± 8.0\\ \cline{2-11} 
B2&\multicolumn{1}{|l|}{U-net-R~\cite{cheukICPR}} & 81.2 ± 6.1& \textbf{61.1 ± 11.1}& \textbf{69.1 ± 8.1}& 51.0 ± 10.3& \textbf{61.3 ± 12.4}& 55.3 ± 10.6& 26.2 ± 9.8& \textbf{31.8 ± 12.4}& 28.6 ± 10.7 \\ \cline{2-11} 
B3&\multicolumn{1}{|l|}{U-net VAT} & 79.1 ± 6.6& 56.9 ± 11.7& 65.4 ± 8.7& 52.4 ± 11.9& 60.1 ± 12.6& 55.5 ± 11.3& 26.2 ± 9.8& 30.3 ± 11.6& 27.8 ± 10.3 \\ \cline{2-11} 
B4&\multicolumn{1}{|l|}{U-net-R VAT} & 79.7 ± 6.1& 59.9 ± 11.0& 67.7 ± 7.8& 57.2 ± 11.9& 61.0 ± 12.0& 58.6 ± 11.1& 29.2 ± 10.3& 31.3 ± 11.3& 30.0 ± 10.5 \\ \cline{2-11} 
B5&\multicolumn{1}{|l|}{U-net-O} & 88.3 ± 6.3& 38.4 ± 10.2& 52.6 ± 9.8& 81.9 ± 7.9& 50.1 ± 11.3& 61.6 ± 9.8& 41.0 ± 10.1& 25.6 ± 10.1& 31.2 ± 10.2 \\ \cline{2-11}
B6&\multicolumn{1}{|l|}{U-net-RO} & 88.0 ± 6.3& 44.9 ± 11.2& 58.5 ± 10.1& 83.4 ± 8.7& 55.8 ± 12.2& 66.3 ± 10.3& 42.8 ± 10.3& 29.2 ± 10.7& 34.4 ± 10.6 \\ \cline{2-11} 
B7&\multicolumn{1}{|l|}{U-net-O VAT} & 89.5 ± 6.6& 41.0 ± 10.5& 55.4 ± 10.0& \textbf{86.8 ± 8.8}& 52.3 ± 12.1& 64.7 ± 10.7& 44.0 ± 10.8& 27.0 ± 10.4& 33.1 ± 10.7 \\ \cline{2-11} 
B8&\multicolumn{1}{|l|}{U-net-RO VAT} & \textbf{90.0 ± 6.2}& 43.9 ± 10.6& 58.2 ± 9.7& 86.2 ± 8.6& 57.1 ± 11.5& \textbf{68.2 ± 10.0}& \textbf{44.6 ± 11.7}& 30.0 ± 11.0& \textbf{35.6 ± 11.3 }\\ \cline{2-11} 
B9&\multicolumn{1}{|l|}{O\&F~\cite{Hawthorne2017OnsetsAF}} & 89.7 ± 5.8& 37.7 ± 10.4& 52.2 ± 10.4& 85.3 ± 8.6& 51.2 ± 13.1& 63.1 ± 11.4& 41.8 ± 10.5& 25.5 ± 10.2& 31.2 ± 10.4  \\ \cline{2-11} 

&\multicolumn{1}{c|}{\textbf{One-shot}} & P& R& F1          & P             & R             & F1            & P             & R             & F1            \\ \cline{2-11} 
C1&\multicolumn{1}{|l|}{U-net~\cite{cheukICPR}}   & 61.9 ± 5.8& 41.8 ± 9.1& 49.2 ± 6.9& 24.7 ± 7.6& 49.6 ± 10.6& 32.2 ± 7.9& 8.8 ± 5.6& 17.0 ± 8.5& 11.3 ± 6.5        \\ \cline{2-11} 
C2&\multicolumn{1}{|l|}{U-net-R~\cite{cheukICPR}} & 74.8 ± 5.9& 41.2 ± 10.4& 52.2 ± 8.5  &    33.6 ± 9.4& 54.5 ± 11.5& 40.7 ± 8.8   & 12.5 ± 7.2& 19.8 ± 10.0& 15.0 ± 8.1 \\ \cline{2-11} 
C3&\multicolumn{1}{|l|}{U-net VAT}                & 75.6 ± 6.6& \textbf{46.1 ± 11.1}& \textbf{56.2 ± 9.0}& 42.7 ± 11.0& 55.2 ± 12.7& 47.1 ± 9.6& 17.1 ± 7.8& 22.1 ± 9.7& 18.9 ± 8.1 \\ \cline{2-11} 
C4&\multicolumn{1}{|l|}{U-net-R VAT}              & 71.6 ± 6.0& 43.6 ± 9.5& 53.3 ± 7.3& 36.4 ± 8.5& \textbf{61.5 ± 12.0}& 45.2 ± 8.5& 13.6 ± 6.6& \textbf{22.8 ± 10.4}& 16.8 ± 7.7 \\ \cline{2-11} 
C5&\multicolumn{1}{|l|}{U-net-O}                  & 87.7 ± 7.2& 17.4 ± 7.0 & 28.4 ± 9.3  &    \textbf{85.7 ± 6.6}& 27.7 ± 10.0& 40.9 ± 11.3  & \textbf{34.5 ± 10.6}& 11.4 ± 6.1& 16.8 ± 7.7 \\ \cline{2-11} 
C6&\multicolumn{1}{|l|}{U-net-RO}                 & 87.7 ± 7.5& 21.4 ± 9.0& 33.4 ± 11.1  & 82.0 ± 7.1& 33.4 ± 11.7& 46.3 ± 11.9& 34.4 ± 10.0& 14.6 ± 7.9& 20.0 ± 9.0 \\ \cline{2-11} 
C7&\multicolumn{1}{|l|}{U-net-O VAT}              & 72.9 ± 9.4& 22.9 ± 8.8 & 33.5 ± 9.1  &    57.9 ± 10.2& 43.6 ± 12.4& 47.9 ± 7.6  & 17.1 ± 7.8& 13.0 ± 7.0& 14.2 ± 6.8 \\ \cline{2-11} 
C8&\multicolumn{1}{|l|}{U-net-RO VAT} & 86.1 ± 6.8& 31.4 ± 9.8& 45.0 ± 10.3& 77.2 ± 9.8& 51.5 ± 12.4& \textbf{60.7 ± 9.6}& 31.4 ± 10.4& 21.1 ± 9.1& \textbf{24.8 ± 9.2} \\ \cline{2-11} 
C9&\multicolumn{1}{|l|}{O\&F~\cite{Hawthorne2017OnsetsAF}}     & \textbf{88.0 ± 6.8}& 12.4 ± 4.8& 21.3 ± 7.2& 83.5 ± 8.2& 30.5 ± 10.4& 43.7 ± 11.2& 23.1 ± 11.4& 8.3 ± 4.7& 11.9 ± 6.3  \\ \cline{2-11} 

\end{tabular}}
 \label{tab:full_MAPS}
\end{table*}

\begin{table*}[h]
\centering
 \caption{Transcription accuracy (mean ± STD) in terms of different metrics and their respective precision (P), recall (R), and F1-score (F1) when training on the different variations of the MusicNet dataset. The accuracy values are averaged over the test clips in the dataset.}
\small 
\resizebox{\textwidth}{!}
{\begin{tabular}{ll|ccc|ccc|ccc|}
\cline{3-11}
&\multicolumn{1}{c|}{}                             & \multicolumn{3}{c|}{\textbf{Frame}} & \multicolumn{3}{c|}{\textbf{Note}}            & \multicolumn{3}{c|}{\textbf{Note w/ offset}}  \\ \cline{3-11} 
&\multicolumn{1}{c|}{\textbf{Strings}}                             & P              & R    & F1          & P             & R             & F1            & P             & R             & F1            \\ \cline{2-11}
D1&\multicolumn{1}{|l|}{U-net-O} & 70.0 ± 8.4& 25.3 ± 13.4& 35.7 ± 13.8& 59.9 ± 10.7& 24.6 ± 13.0& 33.9 ± 13.6& 35.4 ± 11.3& 15.0 ± 9.9& 20.5 ± 11.0 \\ \cline{2-11}
D2&\multicolumn{1}{|l|}{U-net-RO} & 79.1 ± 1.1& 39.3 ± 18.7& 50.1 ± 16.6& 68.5 ± 12.3& 39.4 ± 17.2& 48.6 ± 16.1& 49.0 ± 17.1& 29.9 ± 17.8& 36.3 ± 18.5 \\ \cline{2-11}
D3&\multicolumn{1}{|l|}{U-net-O VAT} & 65.2 ± 18.9& 27.6 ± 13.5& 38.2 ± 15.6& 52.7 ± 14.2& 27.8 ± 14.5& 35.7 ± 14.9& 31.5 ± 13.4& 17.4 ± 12.6& 22.0 ± 13.5 \\ \cline{2-11}
D4&\multicolumn{1}{|l|}{U-net-RO VAT} & 78.0 ± 4.0& 36.9 ± 10.5& 49.5 ± 10.1& \textbf{69.6 ± 13.7}& 37.6 ± 11.4& 48.4 ± 11.9& \textbf{50.1 ± 18.3}& 27.8 ± 13.4& 35.4 ± 15.4 \\ \cline{2-11}
D5&\multicolumn{1}{|l|}{U-net~\cite{cheukICPR}}   & 67.1 ± 6.9& 50.7 ± 12.9& 57.4 ± 10.5& 41.8 ± 12.5& 46.1 ± 12.7& 43.7 ± 12.5& 24.6 ± 13.4& 26.7 ± 14.0& 25.6 ± 13.7        \\ \cline{2-11}
D6&\multicolumn{1}{|l|}{U-net-R~\cite{cheukICPR}} & 71.7 ± 2.9& \textbf{62.8 ± 10.6}& 66.6 ± 6.9& 53.1 ± 10.7& 57.4 ± 11.9& 55.1 ± 11.1& 37.6 ± 14.1& \textbf{40.9 ± 16.4}& 39.1 ± 15.2 \\ \cline{2-11}
D7&\multicolumn{1}{|l|}{U-net VAT} & 76.1 ± 7.2& 52.4 ± 11.9& 61.7 ± 10.2& 58.1 ± 13.9& 51.2 ± 15.1& 54.2 ± 14.3& 39.6 ± 17.8& 35.4 ± 18.4& 37.2 ± 18.1 \\ \cline{2-11}
D8&\multicolumn{1}{|l|}{U-net-R VAT} &\textbf{ 78.9 ± 4.8}& 60.7 ± 9.8& \textbf{68.4 ± 7.7}& 63.6 ± 13.8& \textbf{58.8 ± 14.3}& \textbf{61.0 ± 13.8}& 43.3 ± 18.8& 40.2 ± 18.9& \textbf{41.6 ± 18.7} \\ \cline{2-11}
D9&\multicolumn{1}{|l|}{O\&F~\cite{Hawthorne2017OnsetsAF}} & 75.3 ± 3.1& 22.5 ± 12.0& 33.1 ± 14.6& 69.0 ± 15.8& 22.6 ± 12.1& 32.9 ± 14.8& 41.7 ± 18.2& 15.3 ± 11.2& 21.7 ± 14.5  \\ \cline{2-11}
D10&\multicolumn{1}{|l|}{Multi-Inst~\cite{MultiInstrument}} & 71.4 ± 6.2& 59.5 ± 13.1& 64.5 ± 9.6& 56.5 ± 16.7& 45.1 ± 18.4& 49.9 ± 17.8& 33.7 ± 19.3& 27.8 ± 18.4& 30.3 ± 18.9 \\ \cline{2-11}

&\multicolumn{1}{c|}{\textbf{Woodwinds}}                             & P              & R    & F1          & P             & R             & F1            & P             & R             & F1            \\ \cline{2-11}
E1&\multicolumn{1}{|l|}{U-net~\cite{cheukICPR}}  & 62.6 ± 9.6& 65.7 ± 2.9& 63.5 ± 3.7& 33.6 ± 2.4& 39.9 ± 1.1& 36.4 ± 1.0& 11.6 ± 1.4& 13.7 ± 0.3& 12.5 ± 0.9        \\ \cline{2-11}
E2&\multicolumn{1}{|l|}{U-net-R~\cite{cheukICPR}} & 65.4 ± 6.7& 71.4 ± 4.3& \textbf{67.8 ± 1.7}& 40.5 ± 5.8& \textbf{52.5 ± 0.6}& 45.4 ± 3.5& 16.9 ± 5.2& 21.3 ± 3.5& 18.7 ± 4.7 \\ \cline{2-11}
E3&\multicolumn{1}{|l|}{U-net VAT} & 69.0 ± 9.6& 62.4 ± 1.2& 65.1 ± 3.7& 44.4 ± 3.1& 40.5 ± 2.5& 42.2 ± 0.0& 16.1 ± 1.6& 14.6 ± 0.5& 15.3 ± 0.5 \\ \cline{2-11}
E4&\multicolumn{1}{|l|}{U-net-R VAT} &\textbf{ 69.8 ± 8.3}& 65.8 ± 2.0& 67.4 ± 2.8& \textbf{48.6 ± 3.5}& 47.9 ± 1.1& \textbf{48.2 ± 1.2}& \textbf{22.1 ± 3.1}& \textbf{21.7 ± 1.0}& \textbf{21.8 ± 2.0} \\ \cline{2-11}
E5&\multicolumn{1}{|l|}{Multi-Inst~\cite{MultiInstrument}} & 64.4 ± 9.4&\textbf{ 71.8 ± 2.7}& 67.3 ± 4.1& 43.5 ± 2.8& 33.6 ± 0.6& 37.9 ± 0.7& 17.1 ± 2.5& 13.2 ± 0.9& 14.9 ± 1.5 \\ \cline{2-11}

\end{tabular}}
 \label{tab:string_MusicNet}
\end{table*}
\section{Applications}
\label{sec:application}
Our proposed semi-supervised framework allows for two important applications: continual learning and knowledge transfer to unseen music genres. We will discuss these two properties and their potential applications.

\subsection{Continual Learning}
The loss function of our proposed semi-supervised AMT framework contains a supervised term $L_\text{l}$ and an unsupervised term $L_\text{ul}$. Even when we encounter new unseen, unlabelled data, we can still use this new data to minimize the unsupervised part of the model $L_\text{ul}$. That means, the proposed model can be \emph{retrained} with any new data that was not collected before. Therefore, our proposed framework is capable of improving itself via new unlabelled data.

To confirm our framework's ability of continual learning, we take the string and woodwind subset of MusicNet as an example. We first train our models for 4,000 epochs (row 1 and 4 of Table~\ref{tab:retain}, denoted as ``4k''), and save the weights. These weights are then used as starting weights when we train the model for another 4,000 epochs with two different conditions: (1) without new data (row 2 and 5, denoted as ``8k''); (2) using the test data as the unlabelled data as well as the existing data (row 3 and 6, denoted as ``4k + 4k''). For the string subset of MusicNet, the model has already converged at 4,000 epochs, additional supervised training does not change the performance much. When we include the test data as the unlabelled data, it further pushes the accuracy around 1 percentage point higher. The same goes for the woodwind dataset. Although the improvement is relatively subtle at the moment, we plan to investigate ways to further improve this in future research. This property leads to the next application.

\begin{figure}[ht]
  \centering
  \includegraphics[width=\linewidth]{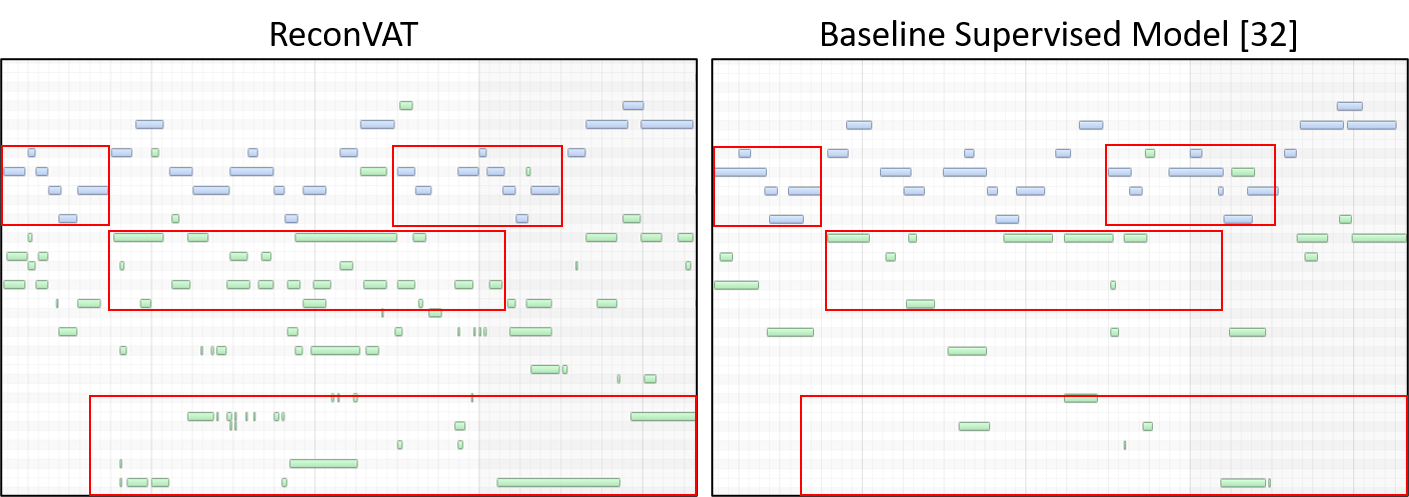}
  \caption{Transcribed piano rolls for J-pop. Blue notes indicate the main melody; green notes indicate the accompaniment (piano and drums). The red boxes indicate major differences in the piano rolls.}
  \Description{Our proposed semi-supervised framework.}
  \label{fig:application}  
\vspace{-4mm}  
\end{figure}

\subsection{Case Study: Transcribing Unseen Genres}
In some cases, we have some labels in one data domain, while the target domain we are interested in might not contain any labels at all. A model that can be trained on one domain and its knowledge then transferred to the target domain will be very useful. For example, we have some labelled data for classical woodwind music, but we want our model to be able to transcribe clarinet covers of Japanese pop music. Our proposed framework, as shown in Figure~\ref{fig:application}, is capable of tackling this task.

We downloaded a few Japanese cover songs from YouTube, and we study the transcription results produced by both the ``woodwind 4k + 4k'' model reported in Table~\ref{tab:retain} and the best supervised baseline model~\cite{MultiInstrument} trained for 8,000 epochs using only labelled data. Due to page limitations, we only show one of the cover songs ``Lemon'' in Figure~\ref{fig:application}. More examples can be found in the demo page provided as part of the supplementary material\footnote{https://kinwaicheuk.github.io/ReconVAT}. Since Japanese Pop music is not included in the woodwind version of MusicNet data, the supervised baseline model trained on only MusicNet produces a piano roll with a lot of missing details such as the melody and the bass indicated by the red boxes. Training the supervised model for more epochs does not help as the loss already converged. Our proposed semi-supervised framework (ReconVAT) trained on both labelled and unlabelled data tries to capture more details than the fully supervised baseline model~\cite{MultiInstrument}. For example, in the upper part of the piano roll (Figure~\ref{fig:application}), the baseline model fails to capture the rhythmic patterns that are only found in pop music. Moreover, there are also rhythmic patterns for the piano accompaniment that are specific to pop music, and the baseline model failed to transcribe these unseen piano patterns (middle and lower part in Figure~\ref{fig:application}). One might argue that the transcription result for ReconVAT is noisier in the bass region (bottom part of the piano roll). This is due to the drum patterns in pop music. Since the labelled data in MusicNet does not contain any drum beat as the accompaniment, the baseline model simply ignores the drum sounds in the pop music. Our proposed model, however, is aware of the presence of the drum beats by training with the unlabelled pop music. It therefore attempts to transcribe the drum beats, making the transcription slightly noisier than the baseline model. Nonetheless, this example shows the success of our proposed model in transcribing unseen music genres. 
\begin{table}[t!]
\centering
 \caption{Continual learning of the AMT model on the String and Woodwind subsets of MusicNet. Our proposed framework (4k+4k) has the ability to adapt to unseen and unlabeled new data to improve itself.}
\small
{\begin{tabular}{l|c|c|c|}
\cline{2-4}
\multicolumn{1}{c|}{}  & \textbf{Frame}              & \textbf{Note}    & \textbf{Note w/ offset}\\ \hline
\multicolumn{1}{|l|}{String 4k} & 68.4 ± 7.7& 61.0 ± 13.8& 41.6 ± 18.7 \\ \hline
\multicolumn{1}{|l|}{String 8k} & 67.7 ± 8.0& 61.1 ± 13.5& 41.5 ± 18.8 \\ \hline
\multicolumn{1}{|l|}{String 4k+4k} & \textbf{68.7 ± 8.0} & \textbf{62.7 ± 13.3}& \textbf{42.8 ± 18.9} \\ \hline
\multicolumn{1}{|l|}{Woodwind 4k} & 67.4 ± 2.8& 48.2 ± 1.2& 21.8 ± 2.0 \\ \hline
\multicolumn{1}{|l|}{Woodwind 8k} & \textbf{68.1 ± 2.8}& 50.9 ± 1.3& 23.3 ± 3.2 \\ \hline
\multicolumn{1}{|l|}{Woodwind 4k+4k} & 66.6 ± 0.4 & \textbf{51.7 ± 2.2}& \textbf{23.9 ± 4.6} \\ \hline

\end{tabular}}
\vspace{-6pt}
 \label{tab:retain}
\end{table}
\section{Discussion}
Although VAT is found to be useful for our proposed ReconVAT, in our pilot study we observed instability of VAT in some cases. More specifically, we observed that VAT does not work well with some of the baseline models. Whenever VAT is used, the transcription accuracy for the baseline model will collapse to zero. Even when we pretrain the baseline model to first reach their best performance, the moment the VAT kicks in, it causes a sudden increase in the transcription loss after only one forward step and weight update. The transcription loss does not decrease when the VAT component is present.  We discovered that the dropout layers are the culprit causing this problem. Removing the dropout layers from the baseline models can prevent the above-mentioned problem from happening, but at the same time, the transcription accuracy for the baseline models are severely comprised without the dropout layers. The dropout layers somehow cause instability of the gradient $g$ (Equation~(\ref{eq:VAT gradient})), making it change a lot during each iteration, and it eventually leads to a vanishing gradient issue, and hence the gradient explosion of the $\frac{g_t}{\norm{g_t}}$ term. Although the models proposed by \citet{Thickstun2017InvariancesAD} and \citet{pedersoli2020improving} do not have any dropout layers, their models are too resources consuming and take too much time to train. Therefore, we do not find enough motivation to apply VAT on them.

The U-net model proposed by~\citet{cheukICPR} and~\citet{Hung2019MusicalCS} does not use any dropout layers, they are compact in size, and work well with our proposed framework. However, we believe that $p(Y_\text{post}|X_\text{spec},\theta)$ may be replaced by any type of model as long as it does not affect the stability of the gradient $g$. Based on these results, we believe that future research opportunities for AMT lie in the semi-supervised or even unsupervised techniques that work well in scenarios with insufficient labelled data instead of just exploring deep fully supervised models that only work well with abundant labelled data.

\section{Conclusion}
In this paper, we proposed a VAT based semi-supervised AMT framework, \textbf{ReconVAT}, that works well for different kinds of musical instruments such as strings and woodwinds. We demonstrated its power of leveraging unlabelled data to enhance the transcription accuracy when the availability of labelled data is limited. Our proposed framework also generalizes better to other genres that are not present in the training dataset such as music covers of Japaneses pop music. The compactness of our model also allows it to be easily deployed in real-world applications\footnotemark[2].

\begin{acks}
This work is supported by A*STAR SING-2018-02-0204, MOE Tier 2 grant MOE2018-T2-2-161, and SRG ISTD 2017 129.
\end{acks}








\bibliographystyle{ACM-Reference-Format}
\balance
\bibliography{sample-base}











\end{document}